\documentclass[preprint,12pt, sort&compress]{elsarticle}

\makeatletter
\def\ps@pprintTitle{%
 \let\@oddhead\@empty
 \let\@evenhead\@empty
 \def\@oddfoot{}%
 \let\@evenfoot\@oddfoot}
\makeatother

\usepackage{amssymb}
\usepackage{amsmath}
\usepackage{array}
\usepackage{multirow}
\usepackage{hyperref}

\usepackage{chngcntr}
\makeatletter

\let\c@table\c@figure
\makeatother 
\counterwithin{figure}{section}
\counterwithin{table}{section}

\begin{document}

\renewcommand{\floatpagefraction}{0.7}

\begin{frontmatter}



\title{Precision study of critical slowing down in lattice simulations of the CP$^{N-1}$ model}


\author{Jonathan Flynn}
\author{Andreas J\"{u}ttner}
\author{Andrew Lawson}
\author{Francesco Sanfilippo}
\address{School of Physics and Astronomy, University of Southampton, Southampton, SO17 1BJ, UK}

\author{}

\address{}

\begin{abstract}

With the aim of studying the relevance and properties of critical slowing down in Monte Carlo simulations of lattice quantum field theories we carried out a high precision numerical study of the discretised two-dimensional CP$^{N-1}$ model at $N=10$ using an over-heat bath algorithm. We identify critical slowing down in terms of slowly-evolving topological modes and present evidence that other observables couple to these slow modes. This coupling is found to reduce however as we increase the physical volume in which we simulate.


\end{abstract}


\end{frontmatter}


\section{Introduction}
\label{Intro}

Two-dimensional CP$^{N-1}$ models are of great interest to lattice QCD practitioners, as they exhibit similar properties to QCD, such as confinement and asymptotic freedom~\cite{1N_expansion_instantons,  Witten_Instantons_1N_Expansion}. Because of their lower dimensionality, they are also much easier and less costly to simulate; for this reason they provide an excellent framework in which to study and test the properties of algorithms. For example, they are very useful for studying the phenomenon of critical slowing down (CSD): an increase in Monte Carlo relaxation time for particular observables when generating a new statistically independent configuration on which that observable can be measured.

Within the context of the CP$^{N-1}$ model, an exponential form of CSD for the topological modes has been well documented~\cite{MCSim_CPN-1_Models, Original_CSD_top_modes, CSD_Topological_modes}. Other observables, such as the magnetic susceptibility, are expected to largely decouple from the topological modes~\cite{Original_CSD_top_modes} and thus the CSD is expected to conform to a simple power law. However potential small deviations from this behaviour have been observed at high values of the correlation length~\cite{HMC_CP9_results}. Our aim is to study CSD in the CP$^{N-1}$ model to a uniquely high level of precision, with the aim of quantifying any deviations from the expected form of scaling for very fine lattices.

Our detailed studies of CSD require the generation of hundreds of millions of lattice configurations; as such we are able to determine certain observables in the CP$^{N-1}$ model to a high level of statistical precision. For this reason it becomes necessary to be alert to numerical errors that algorithms may introduce, which are usually vastly outweighed by statistical errors. We therefore also present the fine-tuning necessary to the over-heat bath algorithm employed to overcome such numerical errors that may become manifest in particularly long Monte Carlo simulations.

The layout of this paper is as follows: in section \ref{sec:Model} we introduce the continuum formulation of the CP$^{N-1}$ model. In section \ref{sec:lat_form} we provide the details of the lattice formulation that we use and detail the observables that we measure in our simulations. In section \ref{sec:overheat} we give an overview of the over-heat bath algorithm that we use in our simulations, as well as detailing the adjustments necessary in order to avoid numerical errors. We then go on to present the results of our numerical simulations. In section \ref{sec:FSS} we first give an overview of the finite volume effects of our model, which motivates the choice of lattice sizes used in our simulations of CSD, which we report in section \ref{sec:CSD}. We demonstrate a deviation from the expected Gaussian scaling for the integrated autocorrelation time of the magnetic susceptibility, which we can clearly attribute to a coupling between the topological modes and other observables. In section \ref{sec:Conclusions} we present our conclusions.

\section{The CP$^{N-1}$ Model}
\label{sec:Model}

The CP$^{N-1}$ model in 2 dimensions is classified by the action~\cite{1N_expansion_instantons, Witten_Instantons_1N_Expansion}:
\begin{align} \label{eq:CPN_action}
	S=\dfrac{1}{g}\int\mathrm{d}^{2}x\left(\overline{D_{\mu}z\left(x\right)}\cdot D^{\mu}z\left(x\right)\right),
\end{align}
 where $z$ is an $N$-component vector of complex scalar fields subject to the constraint
\begin{align}\label{eq:constraint}
	\bar{z}\left(x\right)\cdot z\left(x\right)=1.
\end{align}
 The covariant derivative is defined as
\begin{align}
	D_{\mu}=\partial_{\mu}+iA_{\mu},
\end{align}
 and $A_{\mu}$ is an auxiliary $U\left(1\right)$ gauge field.

Using the composite operator~\cite{MCSim_CPN-1_Models}
\begin{align}
	P\left(x\right)=\bar{z}\left(x\right)\otimes z\left(x\right),
\end{align}
 we can obtain the correlation function 
\begin{align}\label{eq::corr_fct}
	G_{P}\left(x\right)=&\mathrm{Tr}\left\langle P\left(x\right)P\left(0\right)\right\rangle _{conn}=\mathrm{Tr}\left\langle P\left(x\right)P\left(0\right)\right\rangle-\dfrac{1}{N}.
\end{align}
 This function will be of importance for deriving other lattice observables, which we will discuss in the next section.

The main quantity of interest for our CP$^{N-1}$ simulations is the topological charge of the field $z\left(x\right)$, which is defined by~\cite{1N_expansion_instantons, Witten_Instantons_1N_Expansion}:
\begin{align}\label{top_ch_def}
	q\left(x\right)=\dfrac{1}{2\pi}\epsilon_{\mu\nu}\partial^{\mu}A^{\nu}=\dfrac{i}{2\pi}\epsilon_{\mu\nu}\overline{D_{\mu}z\left(x\right)}\cdot D^{\mu}z\left(x\right).
\end{align}
This quantity is related to the topological susceptibility through
\begin{align}
	\chi_{t}=\int\mathrm{d}^{2}x\left\langle q\left(x\right)q\left(0\right)\right\rangle .
\end{align}
 
\section{Lattice Formulation}
\label{sec:lat_form}

The lattice formulation of the CP$^{N-1}$ model that we use follows from employing a first-order discretisation of Eq. (\ref{eq:CPN_action}) using a periodic square lattice of side length $L$. This leads us to the action~\cite{Geometric_Topological_Charge, MCSim_CPN-1_Models, Lattice_form_1, Lattice_CPN_large_N_behaviour}
\begin{align}\label{eq:lattice_action}
	S_{g}=-N\beta\sum_{n,\mu}\left(\bar{z}_{n+\mu}z_{n}\lambda_{n,\mu}+\bar{z}_{n}z_{n+\mu}\bar{\lambda}_{n,\mu}-2\right),
 \end{align}
where we have introduced the gauge links $\lambda_{n,\mu}\in U\left(1\right)$ connecting adjacent lattice sites and $N\beta=1/g$. We index the lattice with with $n=(n_x,n_y)$ and the direction of the link with $\mu=1,2$. While it is possible to integrate out the $U(1)$ gauge field using its equation of motion, it is useful to leave the fields in explicitly. This makes the resulting lattice formulation of the action linear with respect to each variable, which allows the use of local updating algorithms, such as the over heat bath algorithm.

\subsection{Lattice Observables}

We can derive two important observables from the Fourier transform of the correlation function, defined as~\cite{CSD_Topological_modes, MCSim_CPN-1_Models}
\begin{align}\label{eq:FT}
\tilde{G}\left(k\right)=\dfrac{1}{V}\sum_{n,m}\mathrm{Tr}\left\langle P_{n}P_{m}\right\rangle _{conn}\exp\left(i\dfrac{2\pi}{L}\left(n-m\right)\cdot k\right),
\end{align}
where $n$ and $m$ index the points on the lattice, and the 2-momentum of our theory is defined on a periodic lattice by $p=\left(2\pi/L\right)k$. In a finite volume, the entries of $k$ only take integer values in the range $\left[0,N\right)$. Using Eq. (\ref{eq:FT}) we can extract the magnetic susceptibility, $\chi_{m}$, from
\begin{align}
	\chi_{m}=\tilde{G}\left(0,0\right),
\end{align}
and a definition of the correlation length, $\xi_G$, from
\begin{align} \label{eq:corr_len}
	\xi_{G}^{2}=\dfrac{1}{4\sin^{2}\left(\pi/L\right)}\left(\dfrac{\tilde{G}\left(0,0\right)}{\tilde{G}\left(1,0\right)}-1\right).
\end{align}

We will however make use of another definition of the correlation length, $\xi_w$, defined through the wall-wall correlation function, defined as~\cite{MCSim_CPN-1_Models}
\begin{align}
	G_{w}\left(n_{x}-m_{x}\right)=\dfrac{1}{L}\sum_{n_{y},m_{y}}G_{P}\left(n_{x},n_{y};m_{x},m_{y}\right).
\end{align}

On a lattice with periodic boundary conditions, the long distance behaviour of the correlation function takes the form
\begin{align}
	G_{w}\left(x\right)=\dfrac{A_{w}}{2}\exp\left(\dfrac{L}{2\xi_w}\right)\cosh\left(\dfrac{1}{\xi_w}\left(x-\dfrac{L}{2}\right)\right).
\end{align}
We can therefore extract the parameters $A_w$ and $\xi_w$ by fitting the function $G_{w}\left(x\right)$ in a region dominated by the ground state contribution.

We use the geometrical definition of the topological charge, given by~\cite{Geometric_Topological_Charge}
\begin{align}
	q_{n}=\dfrac{1}{2\pi}\mathrm{Im}\left\{ \ln\left[\mathrm{Tr}P_{n+\mu+\nu}P_{n+\mu}P_{n}\right]+\ln\left[\mathrm{Tr}P_{n+\nu}P_{n+\mu+\nu}P_{n}\right]\right\},
\end{align}
with the constraint $\mu\ne\nu$.
 The topological susceptibility of the configuration is then obtained from
\begin{align}
	\chi_{t}=\dfrac{1}{V}\left\langle \left(\sum_{n}q_{n}\right)^{2}\right\rangle .
\end{align}
 
\subsection{Autocorrelation}

In order to quantify the CSD of observables in our simulation, we measure the integrated autocorrelation time~\cite{Automatic_Windowing}. We start by defining the autocorrelation function for an observable $\mathcal{O}$ as:
\begin{align}
\label{eq:auto_func}
	C\left(t\right)=\dfrac{1}{N_{cf}-t}\sum_{n=1}^{N_{cf}-t}\left[\mathcal{O}_{n+t}-\left\langle \mathcal{O}\right\rangle \right]\left[\mathcal{O}_{n}-\left\langle \mathcal{O}\right\rangle \right],
\end{align}
where $N_{cf}$ is the total number of configurations we have in our Monte Carlo sample. The definition of the integrated autocorrelation time for this observable is
\begin{align} \label{eq:int_auto_time}
	\tau_{\mathcal{O}}=\dfrac{1}{2}+\sum_{t=1}^{\infty}\dfrac{C\left(t\right)}{C\left(0\right)}.
\end{align}
 
It is important to take $N_{cf}\gg\tau_{\mathcal{O}}$ in order to obtain an accurate measurement of the integrated autocorrelation time of an observable. For $t>\tau_{\mathcal{O}}$ we begin to capture more noise than signal in our summation, and so in practice we employ a windowing function $\lambda\left(t\right)$ such that $\lambda(t)\simeq1$ for $t<\tau_{\mathcal{O}}$ and $\lambda\left(t\right)\simeq0$ for $t\gg\tau_{\mathcal{O}}$~\cite{Automatic_Windowing}. In our simulations we use the common choice of
\begin{align}
\lambda\left(t\right)=\begin{cases}
1 & \left|t\right|\le M,\\
0 & \left|t\right|>M,
\end{cases}
\end{align}
where the cutoff $M$ is chosen such that $N_{cf}\gg M\ge c\tau_{\mathcal{O}}$. This is referred to as the ``automatic windowing'' algorithm. This method of course introduces a systematic error because of the truncation of the sum. We must therefore choose a large enough window in order to keep this error small. Our primary source of error on the measurement should therefore be statistical, which we can approximate using the estimator~\cite{Automatic_Windowing}:
\begin{align}
	\sigma_{\tau}^{2}\simeq\dfrac{2\left(2M+1\right)}{N_{cf}}\tau_{\mathcal{O}}^{2}.
\end{align}
We also verified that we obtain compatible results using the automatic windowing method of Ref.~\cite{MC_less_errors}.
 
\section{Over-heat bath Algorithm}
\label{sec:overheat}

In order to undertake simulations of the CP$^{N-1}$ model we have implemented an over-heat bath algorithm~\cite{overheatbath}, which is as follows (following Ref.~\cite{MCSim_CPN-1_Models}). We first express the action at a particular lattice point $n$ as a scalar product of two vectors. The local action relating to the fields $z_{n}$ can be written as
\begin{align}
	S_{n,z}=-\beta N\mathrm{Re}\left\{ \bar{z}_{n}\cdot F_{z,n}\right\} ,
\end{align}
 and for the fields $\lambda_{n,\mu}$ as
\begin{align}
	S_{n,\lambda}=-\beta N\mathrm{Re}\left\{ \bar{\lambda}_{n,\mu}F_{\lambda,n,\mu}\right\} .
\end{align}
The explicit forms of the $F$ terms are therefore:
\begin{align}
	F_{z,n}&=2\sum_{\mu}\left(z_{n-\mu}\lambda_{n-\mu,\mu}+z_{n+\mu}\lambda_{n,\mu}\right),\\
	F_{\lambda,n,\mu}&=2z_{n}z_{n+\mu}.
\end{align}
 These expressions can be written in terms of real vectors, $\phi$ and $F_{\phi}$, with $2k$ components (e.g. arranged such that $\phi_{2j}=\mathrm{Re}\left\{ z_{n,j}\right\}$ and $\phi_{2j+1}=\mathrm{Im}\left\{ z_{n,j}\right\}$ for $j=0,...,k-1$). We have $k=N$ and $k=1$ for $z$ and $\lambda$ updates respectively. The contribution of the vector $\phi$ to the action is then given by (dropping the $n$ subscript for readability):
\begin{align}
	S_{\phi}=-\beta N\phi\cdot F_{\phi}=-\beta N\left|F_{\phi}\right|\cos\theta.
\end{align}
To update the vector $\phi$ we simply generate a new angle $\theta_{new}$ from the probability distribution
\begin{align} \label{eq:prob_dist}
	\dfrac{\mathrm{d}p_{k}}{\mathrm{d}\cos\theta}=\left(\sin\theta\right)^{2k-3}\exp\left(\beta N\left|F_{\phi}\right|\cos\theta\right).
\end{align}
The condition for the over-heat-bath algorithm is that the new vector $\phi_{new}$ is chosen by minimising the scalar product between $\phi_{new}$ and $\phi_{old}$, which is satisfied by taking
\begin{align} \label{eq:new_phi}
	\phi_{new}=&\cos\theta_{new}\dfrac{F_{\phi}}{\left|F_{\phi}\right|}-\left(\phi_{old}-\cos\theta_{old}\dfrac{F_{\phi}}{\left|F_{\phi}\right|}\right)\dfrac{\sin\theta_{new}}{\sin\theta_{old}}\\
	=&\cos\theta_{new}\phi_{\parallel}+\sin\theta_{new}\phi_{\perp}.
\end{align}

Our code is vectorized as follows: first we define the $N_{d}$ as the number of doubles that will fit into the vector register of the CPU we employ. We employ red-black ordering as a means to parallelize our code; however we also group adjacent sites of the same colour (i.e. red or black) into blocks of size $N_d$. We then use vector instructions to generate $N_{d}$ new angles to update these $N_{d}$ sites simultaneously. By employing this method of vectorization we must impose the constraint that $L$ must be a multiple of $2N_{d}$. Running on a single Intel Xeon E5-2670 processor, for a single thread we obtain a peak performance of $\sim 5.2$ $\mathrm{GFLOPS}$ for the over-heat bath update procedure alone ($\sim 25\%$ of peak processor performance) and $\sim 5.9$ $\mathrm{GFLOPS}$ for the entire simulation ($\sim 28.5\%$ of peak processor performance). On a lattice of size $L=160$ this corresponds to generating and taking measurements for 188,500 configurations per core hour.

Additionally we remark that for all our runs we took care to ensure that our Monte Carlo time series are thermalised, by verifying that the average of each observable is stable with respect to the number of thermalisation steps.

\subsection{Trial Angle Generation}

\begin{figure}
\begin{center}
	\begin{tabular}{c}
		\includegraphics[width=9cm]{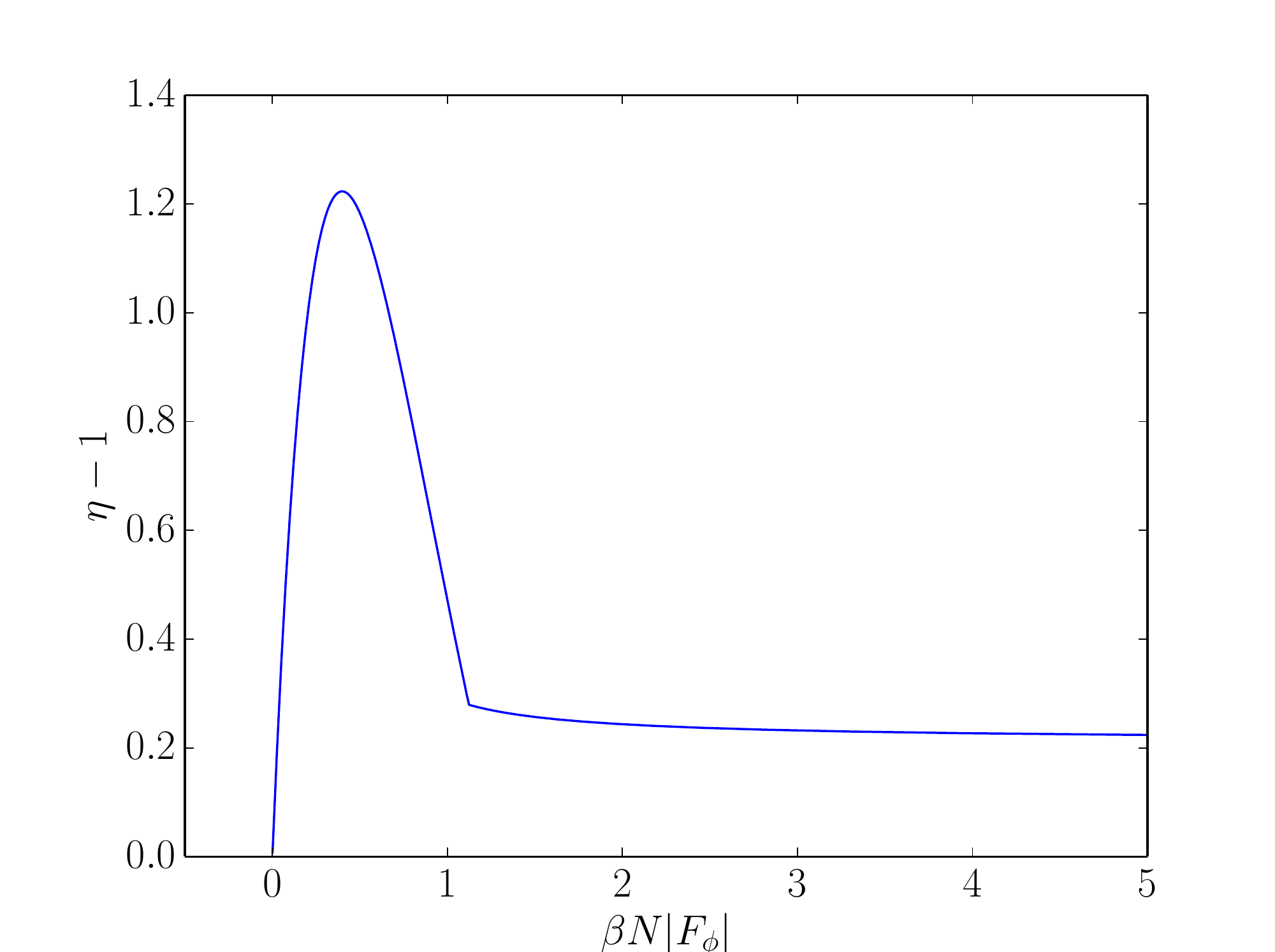} \\
		$(a)$ \\
		\includegraphics[width=9cm]{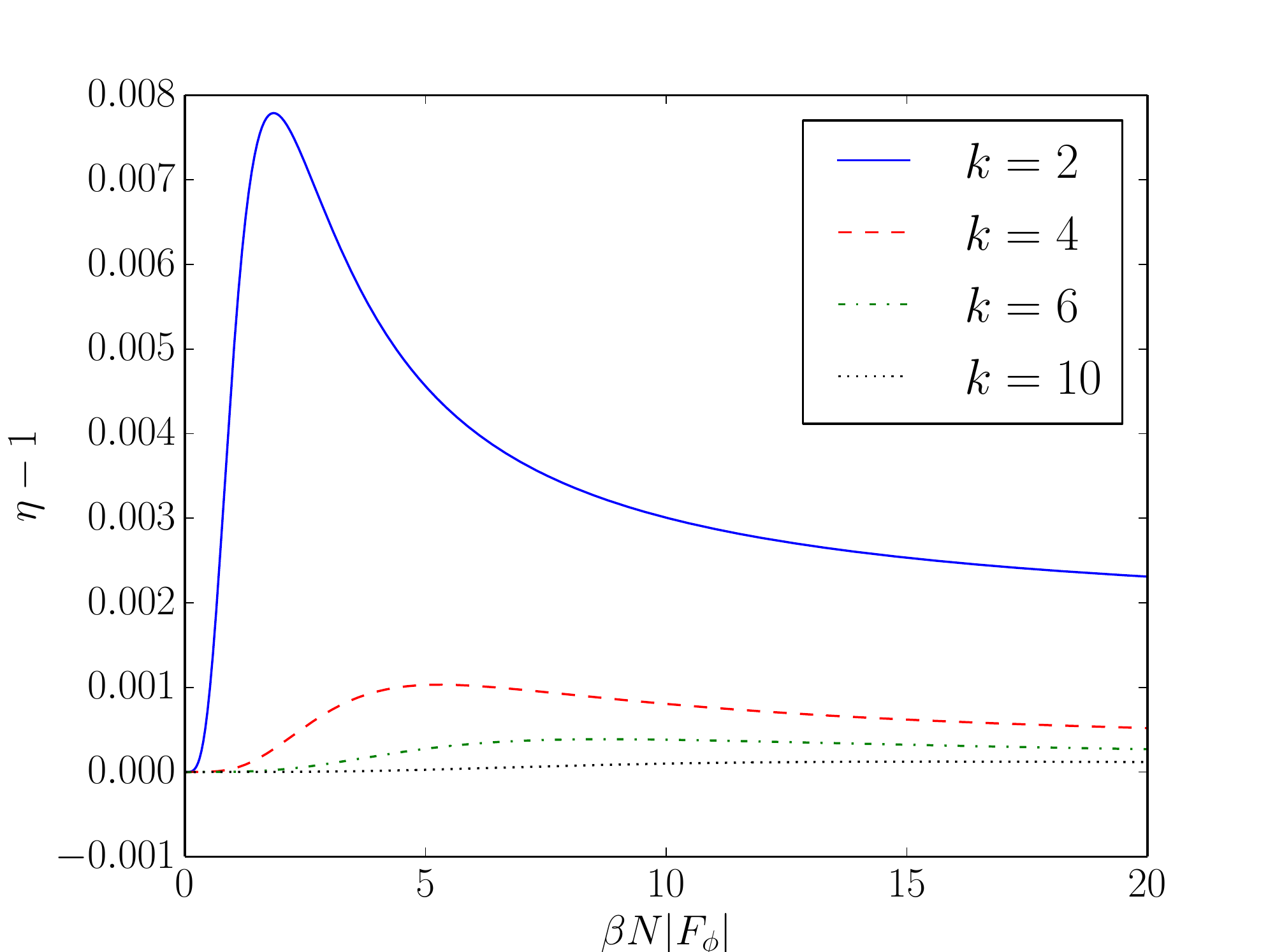} \\
		$(b)$ \\
	\end{tabular}
\end{center}
	\caption{Plots of $\eta-1$ against $\beta N\left|F_{\phi}\right|$ for $(a)$ $k=1$ and $(b)$ $k = 2,4,6,10$.}
	\label{fig:k_vs_eta}
\end{figure}

The details of generating a new angle according to the distribution Eq. (\ref{eq:prob_dist}) are given in the appendix of Ref.~\cite{MCSim_CPN-1_Models}, which we repeat here for convenience. We cannot directly generate angles according to Eq. (\ref{eq:prob_dist}), and so we instead generate angles according to a similar distribution and employ an accept/reject step to ensure that the angles we generate do fit the required distribution. First a trial variable, $\bar{\theta}$, is generated according to the Lorentzian distribution
\begin{align} \label{eq:Lorentz_dist}
\rho_{k}^{t}\left(\theta\right)=\dfrac{1}{1+c^{2}\left(\theta-\theta_{0}\right)^{2}}.
\end{align}
Defining 
\begin{align}
	\zeta=\dfrac{k-1}{\beta N\left|F_{\phi}\right|},
\end{align}
We define the two parameters of this distribution $\theta_{0}$ and $c$ as
\begin{align}
	\theta_{0}=&\arccos\left(\sqrt{1+\zeta^{2}}-\zeta\right),\\
	c=&\sqrt{\beta N\left|F_{\phi}\right|\sqrt{1+\zeta^{2}}}.
\end{align}

The trial variable fitting the distribution in Eq. (\ref{eq:Lorentz_dist}) are obtained from a uniform distribution in $\chi\in\left[0,1\right]$ using
\begin{align}
	\bar{\theta}=\theta_{0}+\dfrac{1}{c}\tan\left[\chi\arctan c\left(\pi-\theta_{0}\right)+\left(\chi-1\right)\arctan c\theta_{0}\right].
\end{align}
To ensure that the trial variables generated match our desired distribution, we accept it with the probability
\begin{align} \label{eq:accept_prob}
	P_{acc}=\dfrac{\rho_{k}\left(\bar{\theta}\right)}{\rho_{k}\left(\theta_{0}\right)}\dfrac{1+c^{2}\left(\bar{\theta}-\theta_{0}\right)^{2}}{\eta},
\end{align}
where $\rho_{k}\left(\theta\right)$ is defined as
\begin{align}
	\rho_{k}\left(\theta\right)=\left(\sin\theta\right)^{2\left(k-1\right)}\exp\left(\beta N\left|F_{\phi}\right|\cos\theta\right),
\end{align}
and the parameter $\eta$ is chosen such that $P_{acc}\leq1$.

In order to choose this free parameter $\eta$, we initially set $\eta=1$ and determine the maximum value that $P_{acc}$ for a range of values of $\beta N\left|F_{\phi}\right|$. The measured values of $P_{acc}^{max}$ then represent the optimal choice for $\eta$.  In Fig. \ref{fig:k_vs_eta} we display our plots of $\eta$ for various values of $k$. For the $z$ updating ($k\ge2$), it is clear that this choice of $\eta$ is only somewhat of a concern for very low $N$. While $\eta=1.01$ would be a safe global choice for all simulations, we choose $\eta=1.0002$ for our CP$^9$ simulations, giving an acceptance rate of $\sim62\%$.

For $\lambda$ updates (where $k=1$) the behaviour of $\eta$ is qualitively very different. For $\beta N\left|F_{\phi}\right|>1.122$ we can analytically determine the best value for $\eta$; to the left of the cusp in Fig.~\ref{fig:k_vs_eta} $(a)$ the maximum of Eq. (\ref{eq:accept_prob}) occurs at the $\theta=\pi$ boundary. To the right the maximum cannot be analytically determined and hence we simply settled for a flat value of $\eta=1.28$, although this choice is of course slightly sub-optimal. Using this method we obtain an acceptance rate of $\sim67\%$. We also tested the possibility of using the "optimized cosh" method for generating random $U(1)$ numbers for the $\lambda$ updates~\cite{Optimised_cosh}. We found that in general this method had an acceptance rate of $\sim90\%$; however because of the increased numerical cost of the latter method, the former gave us a marginally better overall performance.

\subsection{Numerical Errors}

Another issue with the over-heat bath algorithm arises occasionally in the calculation of $\theta_{old}$ from the angle between our $\phi$ and $F_{\phi}$ vectors. From the scalar product of the two vectors we can calculate $\cos\theta_{old}$ to high precision; however when  $\cos\theta_{old}$ is close to $1$, the resultant angle $\theta_{old}$ and ultimately $\sin\theta_{old}$ cannot be determined to as many significant digits because of floating point rounding errors. Therefore extreme values of $\theta_{old}$ may be affected by these numerical errors. Proceeding with the calculation of $\phi_{new}$, the second term in Eq. (\ref{eq:new_phi}) is then likely to introduce a (potentially significant) numerical error. In some rare cases $\theta_{old}$ is evaluated to be zero, which would lead to a hard failure were the algorithm not modified. We therefore implement the following changes to ensure that the algorithm satisfies detailed balance to a higher level of numerical accuracy.

Our solution to the problem is to 'realign' the vector
\begin{align}
	\label{eq:perp_vec}
		\phi_{\perp}=\left(\phi_{old}-\cos\theta_{old}\dfrac{F_{\phi}}{\left|F_{\phi}\right|}\right)\dfrac{1}{\sin\theta_{old}}
\end{align}
 when we measure $\theta_{old}<0.001$ by subtracting off the numerical error. To do this, we take the scalar product of this vector with $F_{\phi}/\left|F_{\phi}\right|$ to determine the numerical deviation, i.e.:
\begin{align}
	\label{eq:num_deviation}
	\dfrac{\phi_{\perp}\cdot F_{\phi}}{\left|F_{\phi}\right|}=\left|\phi_{\perp}\right|\cos\theta_{err}.
\end{align}
We then determine a new perpendicular to $F_{\phi}$ by finding:
\begin{align}
	\phi_{\perp}^{new}=\phi_{\perp} - \left|\phi_{\perp}\right|\cos\theta_{err}\dfrac{F_{\phi}}{\left|F_{\phi}\right|}
\end{align}

Finally we must renormalise this vector to have unit norm. Through our simulations we monitored that this method does indeed produce a vector perpendicular to $F_{\phi}$ up to double precision.

We still have an issue with the above method when we measure $\cos\theta_{old}$ to be 1 within machine precision. In this case we determine $\sin\theta_{old}=0$, and so our attempt at constructing a vector perpendicular to $F_{\phi}$ would fail.  The over-relaxation term vanishes in the limit $\theta_{old}\rightarrow0$; consequently the update will either have no effect or change the sign of $\phi$ (on the condition we normalise $\phi$ to maintain $\bar{z}z=1$). There is zero probability of the update moving the vector out of this (anti-)aligned state and thus detailed balance is broken. We solve this issue by generating a new random unit vector that is perpendicular to $F_{\phi}$ as our $\phi_{\perp}$ term when choosing the remaining degrees of freedom in $\phi$.

\section{Finite Volume Effects}
\label{sec:FSS}

In order to study CSD we measure the integrated autocorrelation times of our observables on a series of lattices scaling towards the continuum (by increasing $\beta$), while holding the physical volume constant (keeping $L/\xi_G$ fixed). Existing studies into the finite volume effects in the CP$^{9}$ model show that $L/\xi_G\gtrsim10$ is a good choice for obtaining sub-percent finite volume effects~\cite{Finite_size_scaling}; however given the high statistical accuracy necessary for our simulations we must necessarily extend these studies to understand the finite volume effects to a greater level of precision.

\begin{figure}
\begin{center}
	\includegraphics[width=9cm]{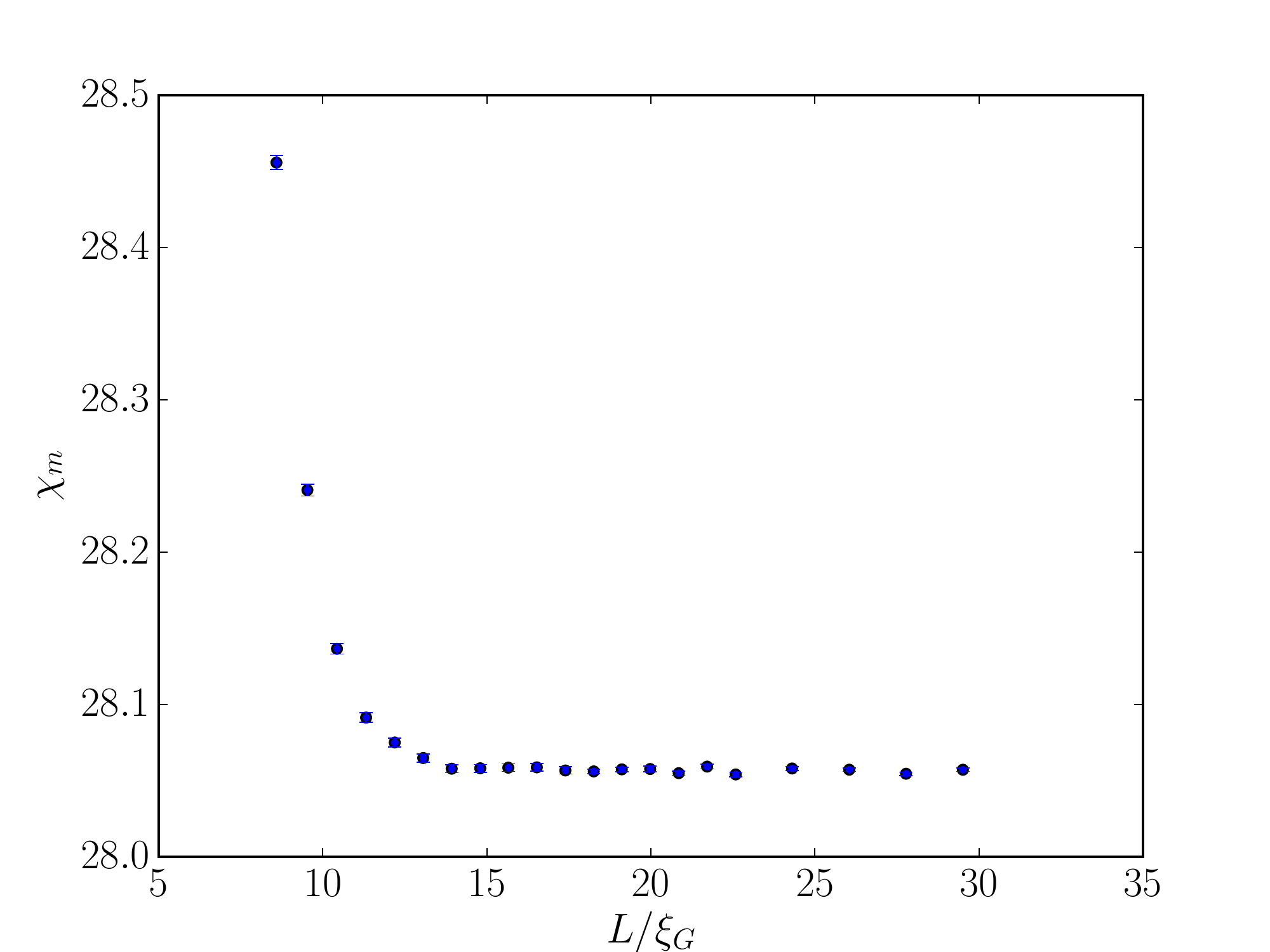} \\
\end{center}
	\caption{Finite volume scaling for $\chi_{m}$ in the CP$^9$ model at $\beta=0.8$.}
	\label{fig:finite_scaling_chi_m}
\end{figure}

\begin{figure}
\begin{center}
	\includegraphics[width=9cm]{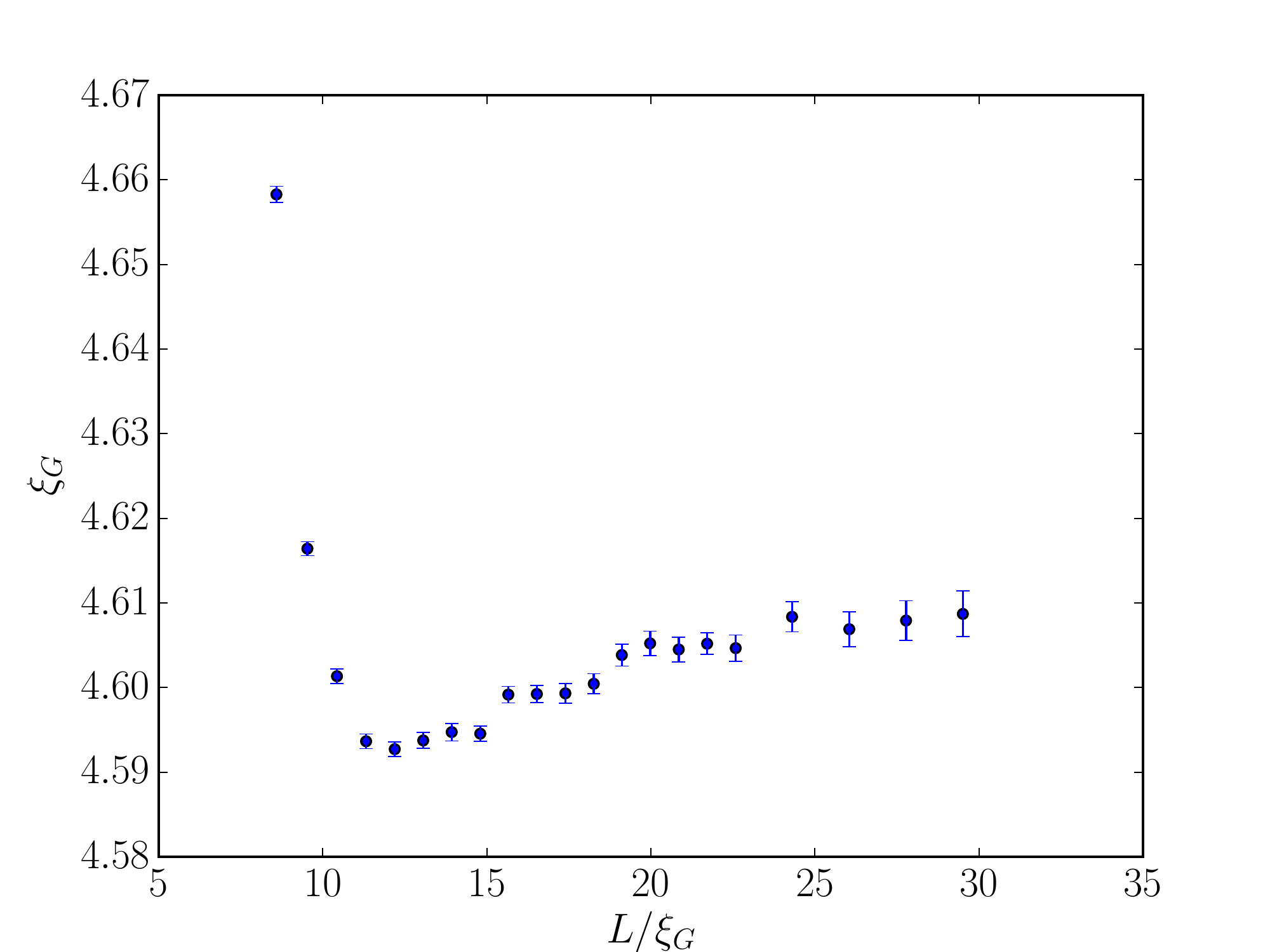} \\
\end{center}
	\caption{Finite volume scaling for $\xi_{G}$ in the CP$^9$ model at $\beta=0.8$.}
	\label{fig:finite_scaling_xi_G}
\end{figure}

\begin{figure}
\begin{center}
	\includegraphics[width=9cm]{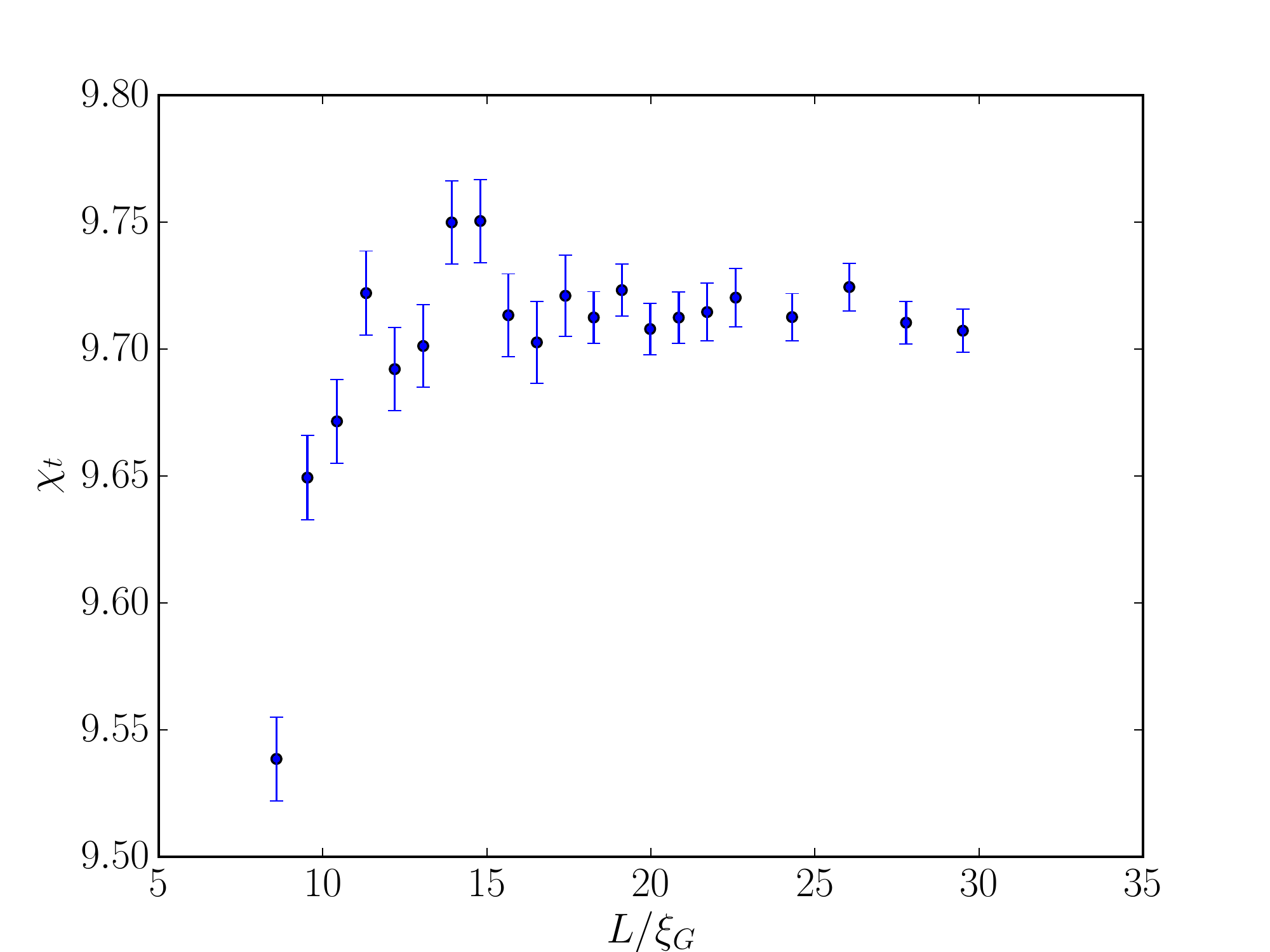}
\end{center}
	\caption{Finite volume scaling for $\chi_{t}$ in the CP$^9$ model at $\beta=0.8$.}
	\label{fig:finite_scaling_chi_t}
\end{figure}

In Figs.~\ref{fig:finite_scaling_chi_m},~\ref{fig:finite_scaling_xi_G} and ~\ref{fig:finite_scaling_chi_t} we present the results of our finite volume analysis for the CP$^9$ model with $\beta=0.8$ for $\chi_m$, $\xi_G$ and $\chi_t$ respectively. We simulated various values of $L$, using 50 million configurations for the smallest values of $L$, ranging up to 140 million configurations for the largest values in order to obtain good resolution of finite volume scaling behaviour. For our studies we would like to choose our parameters such that finite volume effects are under control. Each of our finite volume scaling plots appears to converge to an asymptote within statistical errors (as one would expect from the $1/N$ expansion~\cite{Finite_size_scaling}); it appears that finite volume effects for $\xi_G$ become unnoticable only for $L/\xi_{G}\gtrsim25$ with our level of statistics. Interestingly, the scaling of $\xi_G$ appears to show a turning point around $\xi_G\simeq12$ that is not predicted by the first two terms of the large-$N$ expansion. Finite volume effects for all other observables appear to be negligible within statistical errors from $L/\xi_{G}\simeq15$.

\begin{figure}
\begin{center}
	\includegraphics[width=9cm]{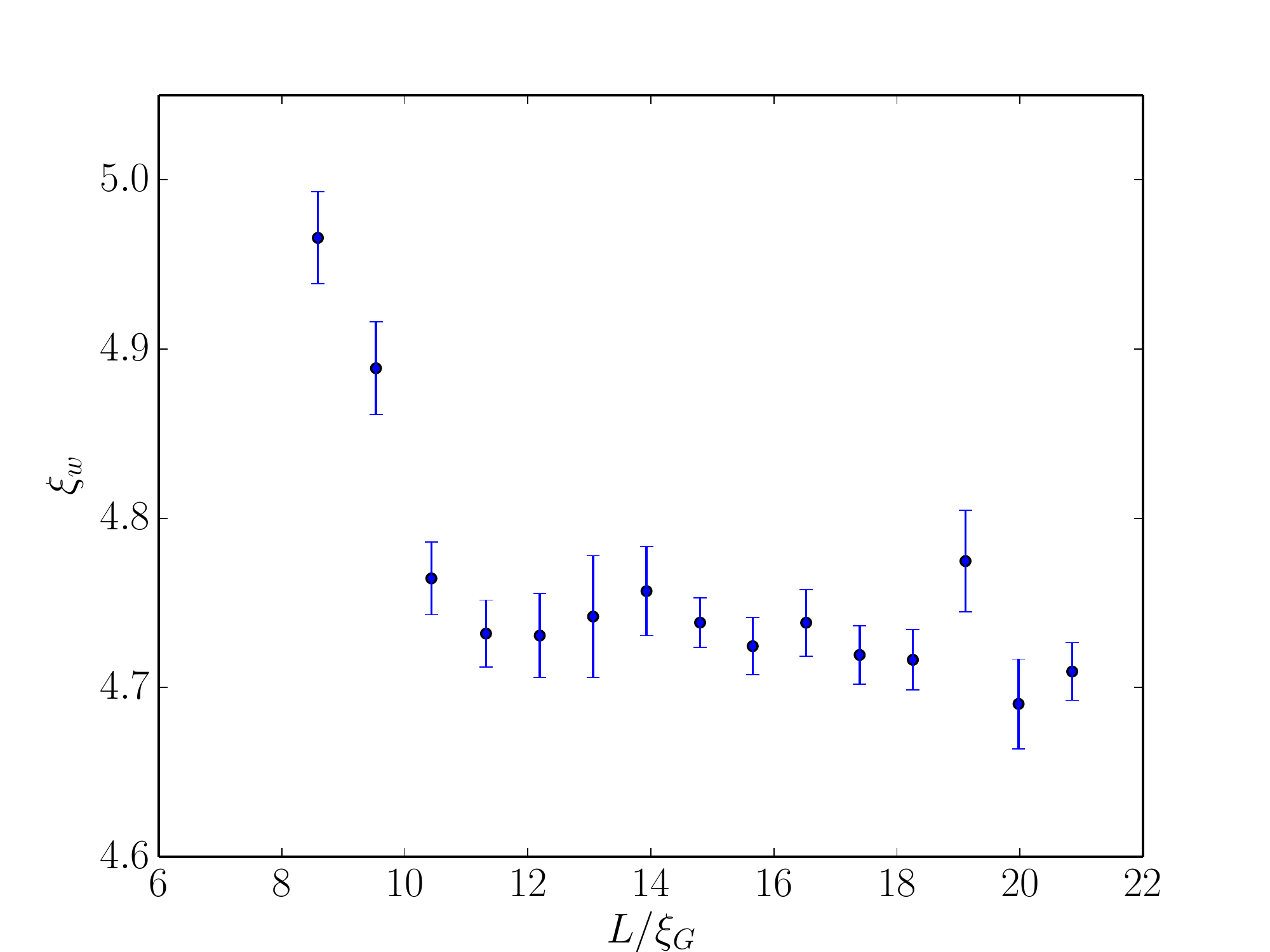}
\end{center}
	\caption{Finite volume scaling for $\xi_{w}$ in the CP$^9$ model at $\beta=0.8$.}
	\label{fig:finite_scaling_xi_w}
\end{figure}

\begin{figure}
	\begin{center}
		\begin{tabular}{c}
			\includegraphics[width=9cm]{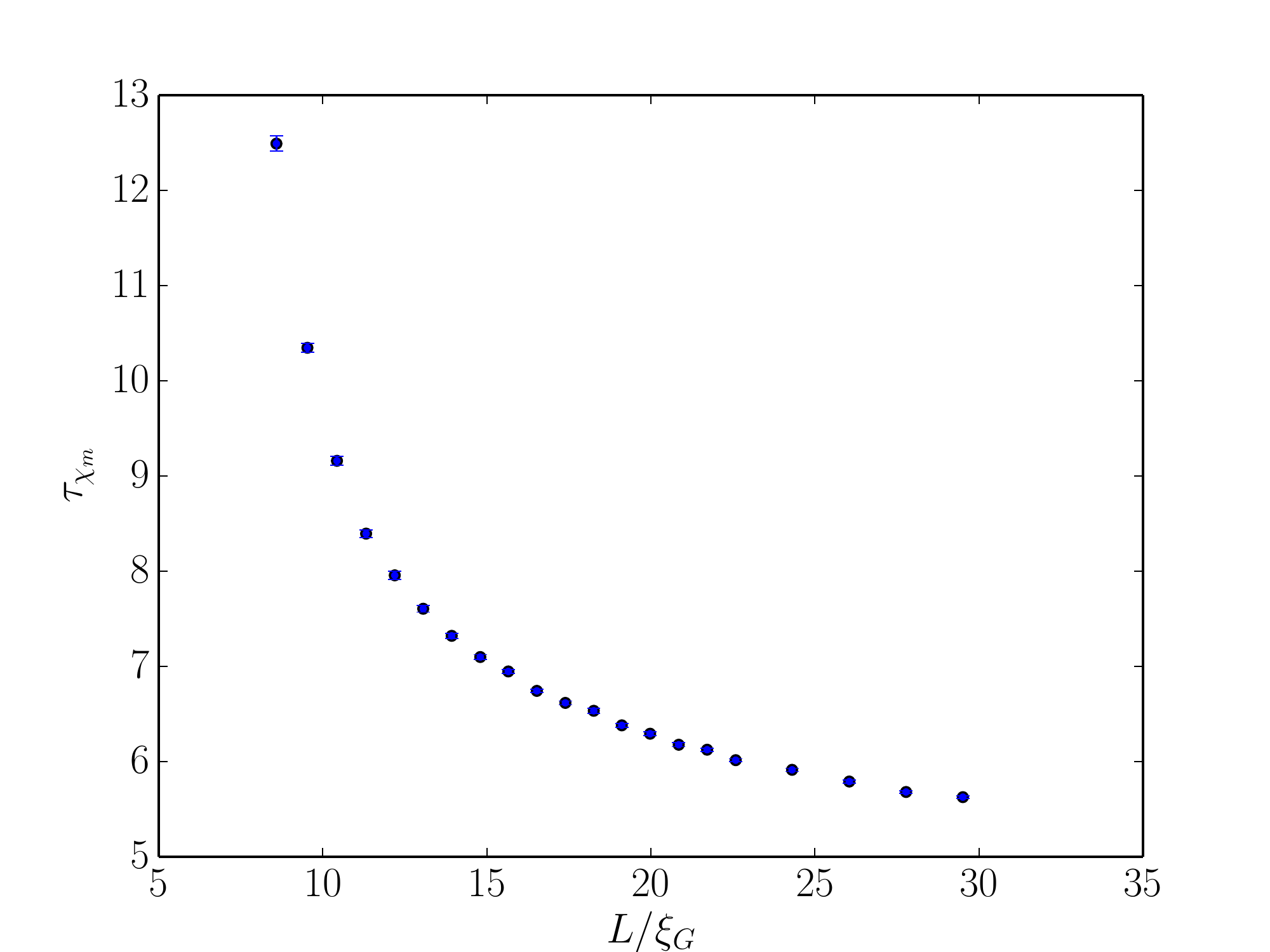} \\
			$(a)$ \\
			\includegraphics[width=9cm]{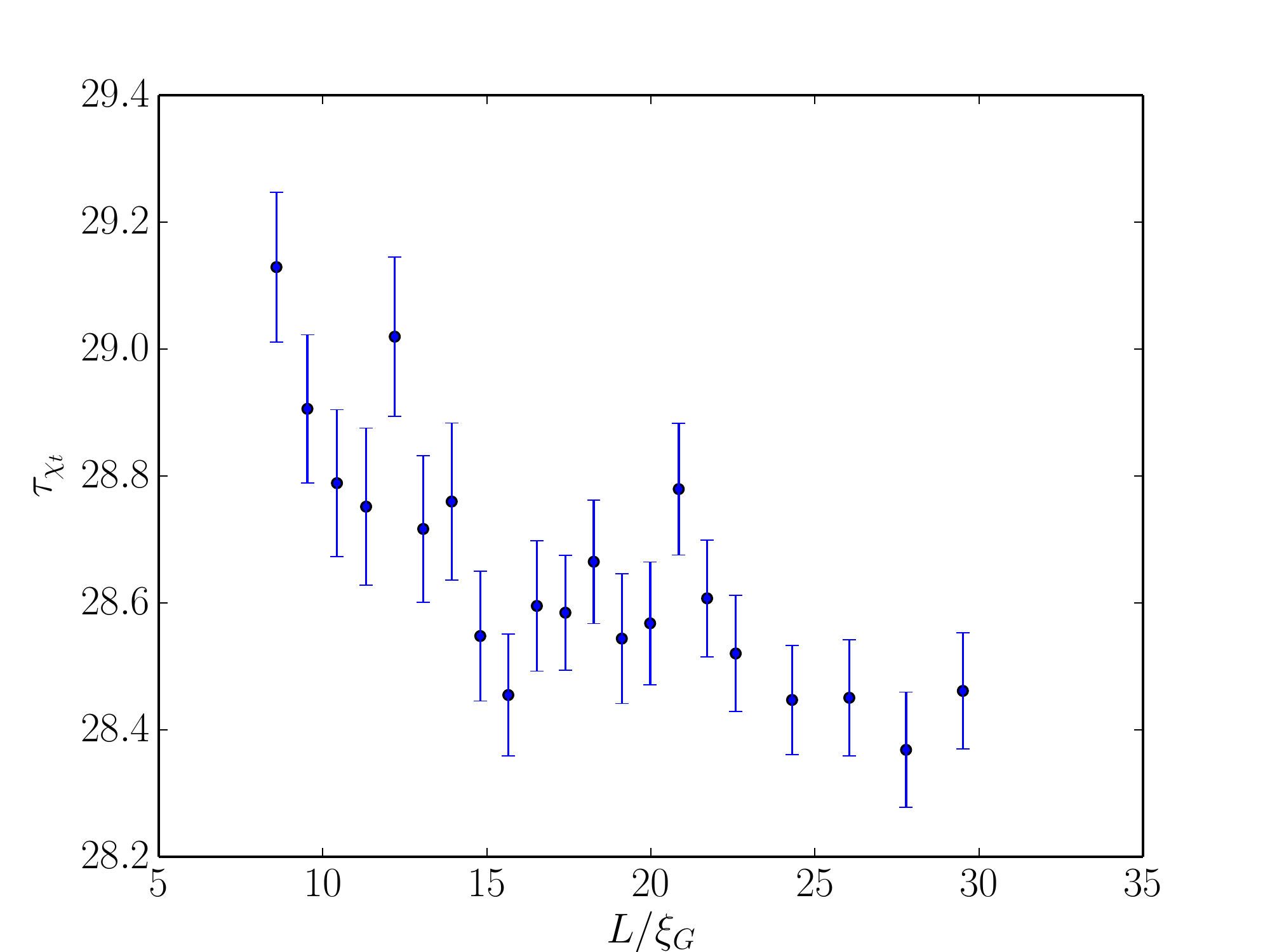} \\
			$(b)$ \\
		\end{tabular}
	\end{center}
	\caption{Finite volume scaling for $(a)$ $\tau_{\chi_{m}}$ and $(b)$ $\tau_{\chi_t}$ in the CP$^9$ model at $\beta=0.8$.}
	\label{fig:finite_scaling_tau}
\end{figure}

 In Fig. \ref{fig:finite_scaling_xi_w} we show the finite volume scaling of $\xi_w$. While the finite volume effects on this definition of the correlation length appear to be more under control, the statistical error on the quantity is very large in comparison to those on $\xi_G$. It is more accurate therefore to simply use $\xi_G$ given that we hold the ratio $L/\xi_G$ approximately constant. Where appropriate we can use an interpolating function to compensate for mistunings in the physical volume $L/\xi_G$. We remark that the finite volume scaling behaviour is independent of UV effects and thus these results hold for higher values of $\beta$. Lastly we note that $\xi_G$ does not reproduce the inverse mass gap of the theory in the continuum limit. However in the scaling region it should scale proportionally to $\xi_w$ and thus it is valid to use $\xi_G$ as the definition of correlation length when we test the scaling of integrated autocorrelation time for our CSD studies.

Finally we consider finite volume effects in the determination of $\tau$ itself. In Fig.~\ref{fig:finite_scaling_tau} we present the finite volume scaling of $(a)$ $\tau_{\chi_m}$ and $(b)$ $\tau_{\chi_t}$. The integrated autocorrelation time itself should vanish in the limit $L/\xi_G\rightarrow\infty$; however again we use an interpolating function to compensate for any slight mistuning of $L/\xi_G$.

\section{Critical Slowing Down}
\label{sec:CSD}

\begin{figure}
\begin{center}
	\includegraphics[width=9cm]{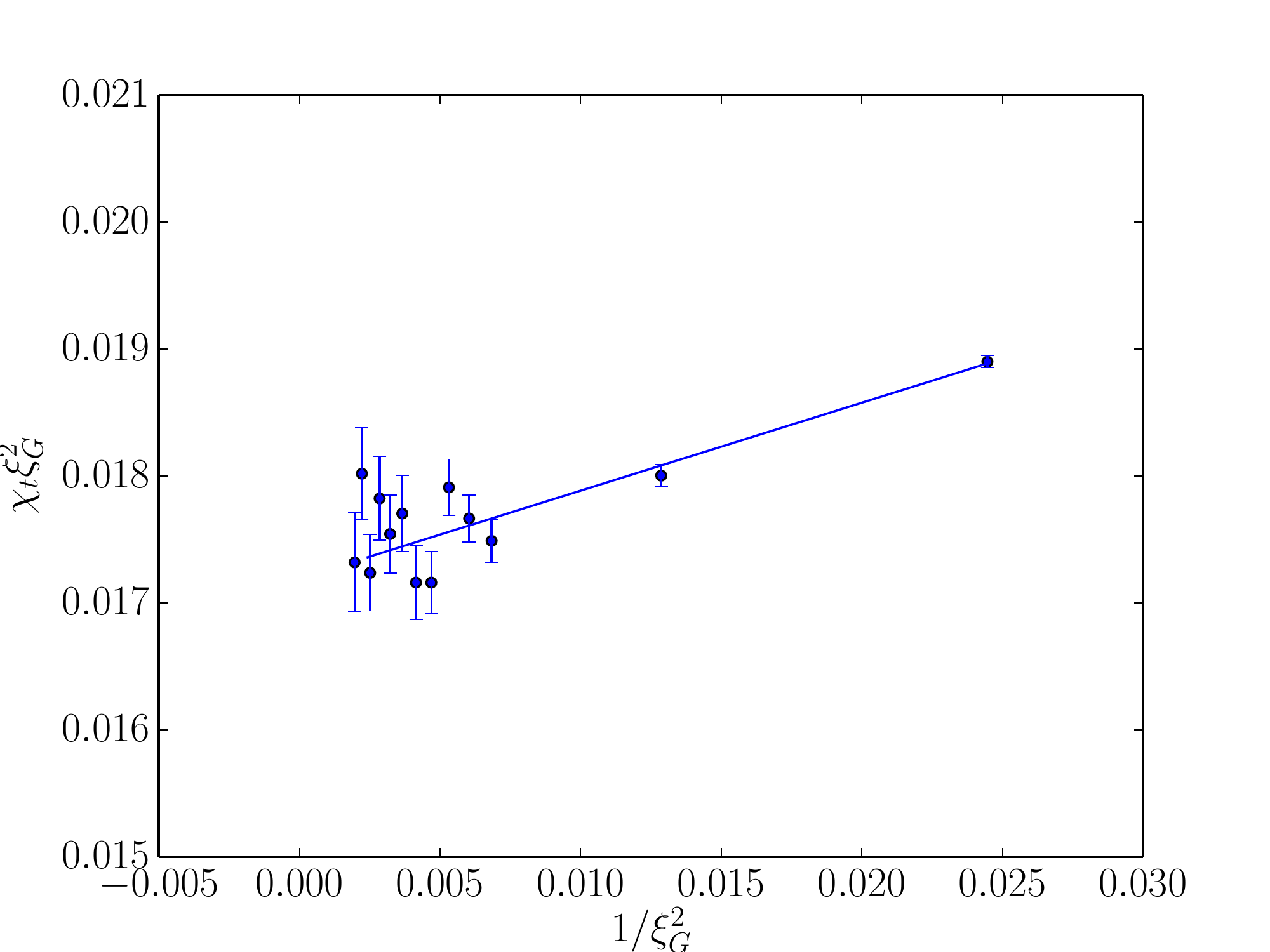} \\
	\vspace{-20pt}
\end{center}
	\caption{Continuum scaling of the dimensionless quantity $\chi_{t}\xi_{G}^{2}$.}
	\label{fig:scaling_region}
\end{figure}

\begin{table}
	\begin{center}
	\begin{tabular}{l l l l l}
		\hline
	   	\multicolumn{1}{c}{$\beta$} & \multicolumn{1}{c}{$\tau_{\chi_m}$} & \multicolumn{1}{c}{$\tau_{\chi_m}^{\mathrm{rem}}$} & \multicolumn{1}{c}{$\tau_{\chi_m}^{\mathrm{slow}}$} & \multicolumn{1}{c}{$\tau_{\chi_t}$} \\
		\hline
		0.8 & \phantom{00}6.95(2) & \multicolumn{1}{c}{-} & \multicolumn{1}{c}{-} & \phantom{000}28.44(9) \\
		0.85 & \phantom{0}10.25(5) & \multicolumn{1}{c}{-} &\multicolumn{1}{c}{-} & \phantom{00}107.6(7) \\
		0.9 & \phantom{0}15.26(12) & \multicolumn{1}{c}{-} & \multicolumn{1}{c}{-} & \phantom{00}452(6) \\
		0.95 & \phantom{0}26.86(58) & 20.58(34) & \phantom{00}6.20(54) & \phantom{0}2338(60) \\
		0.96 & \phantom{0}30.48(62) & 22.58(33) & \phantom{00}7.37(56) & \phantom{0}3173(86) \\
		0.97 & \phantom{0}35.30(89) & 24.74(14) & \phantom{0}10.5(1.3) & \phantom{0}4490(150) \\
		0.98 & \phantom{0}39.1(1.2) & 26.54(30) & \phantom{0}11.75(91) & \phantom{0}6120(230) \\
		0.99 & \phantom{0}48.0(1.6) & 29.18(44) & \phantom{0}18.9(1.1) & \phantom{0}9230(380) \\
		1.0 & \phantom{0}61.8(2.1) & 31.96(34) & \phantom{0}30.5(1.9) & 11980(550) \\
		1.01 & \phantom{0}71.1(2.7) & 35.27(22) & \phantom{0}36.5(4.1) & 17220(850) \\
		1.02 & \phantom{0}97.8(3.4) & 38.02(25) & \phantom{0}62.7(4.7) & 27300(1400) \\
		1.03 & 120.1(4.8) & 41.80(22) & \phantom{0}77.5(6.3) & 36800(1800) \\
		1.04 & 180.6(7.4) & 45.54(26) & 139.6(9.0) & 56500(3200) \\
		1.05 & 203.8(9.2) & 49.88(34) & 158(12) & 70900(4500) \\
		\hline
	\end{tabular}
\end{center}
\caption{Summary of integrated autocorrelation times for our measured observables. For $\beta\le0.9$ the slow mode contribution to $\chi_m$ could not be isolated.}
\label{tab:auto_results}
\end{table}

\begin{table*}
\makebox[\textwidth][c]{
	\begin{tabular}{l l l l l l l l}
		\hline
		\multicolumn{1}{c}{$L$} & \multicolumn{1}{c}{$\beta$} & \multicolumn{1}{c}{Stat} & \multicolumn{1}{c}{$E$} & \multicolumn{1}{c}{$\xi_G$} & \multicolumn{1}{c}{$\xi_w$} & \multicolumn{1}{c}{$\chi_{m}$} & \multicolumn{1}{c}{$10^{5}\chi_{t}$} \\
		\hline
		\phantom{0}72 & 0.8 & \phantom{0}80M & 0.6670232(7) & \phantom{0}4.5992(12) & \phantom{0}4.718(20) & \phantom{0}28.0595(18) & 97.03(11)  \\
		\phantom{0}96 & 0.85 & \phantom{0}80M & 0.6222715(5) & \phantom{0}6.3926(20) & \phantom{0}6.60(3) & \phantom{0}46.863(4) & 46.24(11)\\
		136 & 0.9 & \phantom{0}80M & 0.5838365(3) & \phantom{0}8.815(4) & \phantom{0}9.07(5) & \phantom{0}78.202(8) & 23.17(11) \\
		184 & 0.95 & 100M & 0.55026689(20) & 12.095(6) & 12.40(6) & 130.707(15) & 11.96(11) \\
		192 & 0.96 & 120M & 0.54404507(17) & 12.869(6) & 13.15(4) & 144.880(16) & 10.66(11) \\
		208 & 0.97 & 120M & 0.53797267(16) & 13.709(7) & 14.07(5) & 160.601(20) & \phantom{0}9.53(12) \\
		224 & 0.98 & 120M & 0.53204234(15) & 14.597(8) & 15.01(7) & 178.13(2) & \phantom{0}8.05(12) \\
		232 & 0.99 & 120M & 0.52624990(15) & 15.526(9) & 15.97(8) & 197.48(3) & \phantom{0}7.12(13) \\
		248 & 1.0 & 160M & 0.52058951(13) & 16.528(9) & 16.87(7) & 219.02(3) & \phantom{0}6.48(11) \\
		264 & 1.01 & 200M & 0.51505639(11) & 17.593(10) & 18.15(10) & 242.96(3) & \phantom{0}5.67(10) \\
		288 & 1.02 & 300M & 0.50964608(9) & 18.721(10) & 19.28(7) & 269.50(4) & \phantom{0}5.09(10)  \\
		304 & 1.03 & 450M & 0.50435393(8) & 19.934(10) & 20.46(6) & 299.10(4) & \phantom{0}4.34(8) \\
		320 & 1.04 & 500M & 0.49917929(8) & 21.206(12) & 21.71(6) & 331.83(4) & \phantom{0}4.01(8) \\
		344 & 1.05 & 500M & 0.49410872(8) & 22.558(13) & 23.19(9) & 368.33(5) & \phantom{0}3.40(9) \\
		\hline
	\end{tabular}}
\caption{Results of our Monte Carlo simulations of the CP$^9$ model. Errors were computed using a jackknife analysis.}
\label{tab:sim_results}
\end{table*}

In order to study CSD we performed extensive simulations over a wide range of $\beta$, with $L$ tuned to give $L/\xi_G\simeq 15$. This choice ensures that to our level of statistical precision we can essentially ignore most finite volume effects in the measurements of our observables caused by small fluctations in the quantity $L/\xi_G$. The numerical results of our simulations are given in Table \ref{tab:sim_results}. 

Before we discuss our results of CSD, we can first use our measurements to make a continuum extrapolation of the quantity $\chi_{t}\xi^{2}_{G}$. The action Eq. (\ref{eq:lattice_action}) we employ is correct up to $\mathcal{O}(a^{2})$ cutoff effects, which is corroborated by the apparent linear trend in our scaling plot of $\chi_{t}\xi^{2}_{G}$ in Fig.~\ref{fig:scaling_region}. We can use the plot to make a continuum extrapolation of this quantity, to obtain
\begin{align}
	\chi_{t}\xi^{2}_{G}=0.01719(10)(3).
\end{align}
The central value and first (statistical) error is obtained from fitting the $a^2$ correction on $\xi_G\ge6.6$. The second is a systematic error, which we quote as the change in the continuum value when we fit our full data set while accounting also for an $a^4$ term. The value we obtain is in agreement with previously measured values for this quantity (e.g. see ~\cite{MCSim_CPN-1_Models, Original_CSD_top_modes, CSD_Topological_modes}), which exhibits a significant difference from the large-$N$ expansion result~\cite{1N_expansion_chi_t},
\begin{align}
	\chi_{t}\xi_{G}^{2}=\dfrac{1}{2\pi N}-\dfrac{0.06}{N^{2}}+\mathcal{O}\left(\dfrac{1}{N^3}\right)\stackrel{N=10}{\simeq}0.153.
\end{align}

\begin{figure}
\begin{center}
	\includegraphics[width=9cm]{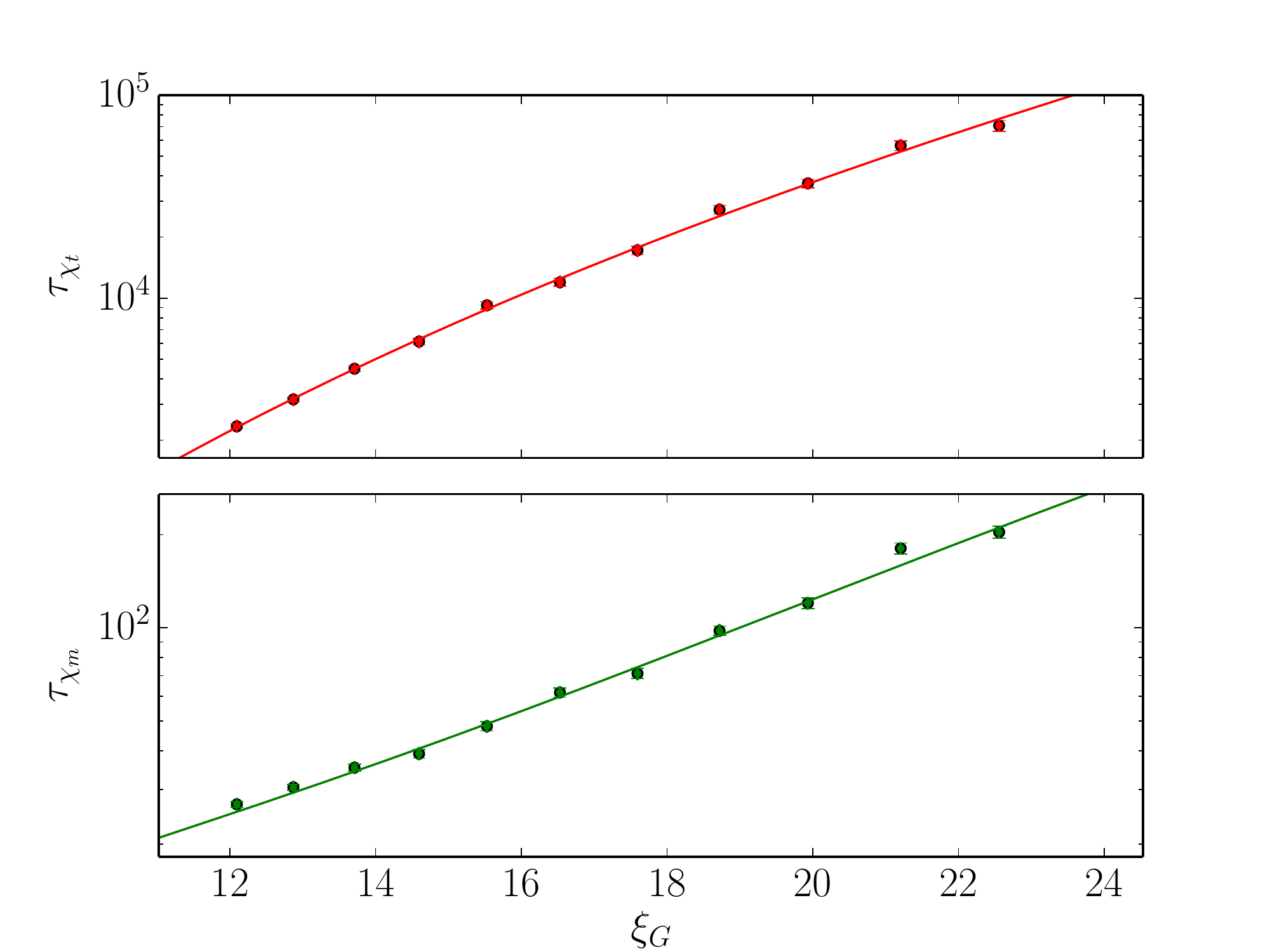} \\
	\vspace{-20pt}
\end{center}
	\caption{Plots of the scaling of $\tau_{\chi_m}$ (green, below) and $\tau_{\chi_t}$ (red, above) with $\xi_G$.}
	\label{fig:auto_results}
\end{figure}

To quantify the CSD of our observables we fit the scaling of the integrated autocorrelation time with $\xi_G$ to the expected relations~\cite{CSD_Topological_modes}, i.e. power law scaling for quasi-Gaussian modes,
\begin{equation}
	\label{eq:gaussian_scaling}
	\tau\sim a\xi^b,
\end{equation}
and exponential scaling for topological modes,
\begin{equation}
	\label{eq:top_scaling}
	\tau\sim a\exp\left(b\xi^{c}\right).
\end{equation}
In general we expect $b\simeq2$ for quasi-Gaussian modes, although using our over-relaxation algorithm we may observe $1\le b<2$ \cite{Overrelation_algos}.

In Fig.~\ref{fig:auto_results} we display the results of our autocorrelation analysis, where we have used an automatic windowing procedure to determine the integrated autocorrelation times. The exponential relation for the topological modes fits very well in the region $\xi_G>12.09$, with $\chi^2/\mathrm{dof}\simeq1.03$ and with fitted parameters $a=0.1(1.6)\times10^{-5}$, $b=12(10)$, $c=0.24(13)$. For comparison, power law scaling for topological modes is also a reasonable fit, albeit with a higher value of $\chi^2/\mathrm{dof}\simeq1.31$, with fitted parameters $a=2.2(4)\times10^{-3}$, $b=5.56(6)$. The $p$-values for these fits are $p=0.41$ and $p=0.23$ respectively, and thus either model gives a valid description of the data.

\begin{figure}
\begin{center}
	\includegraphics[width=9cm]{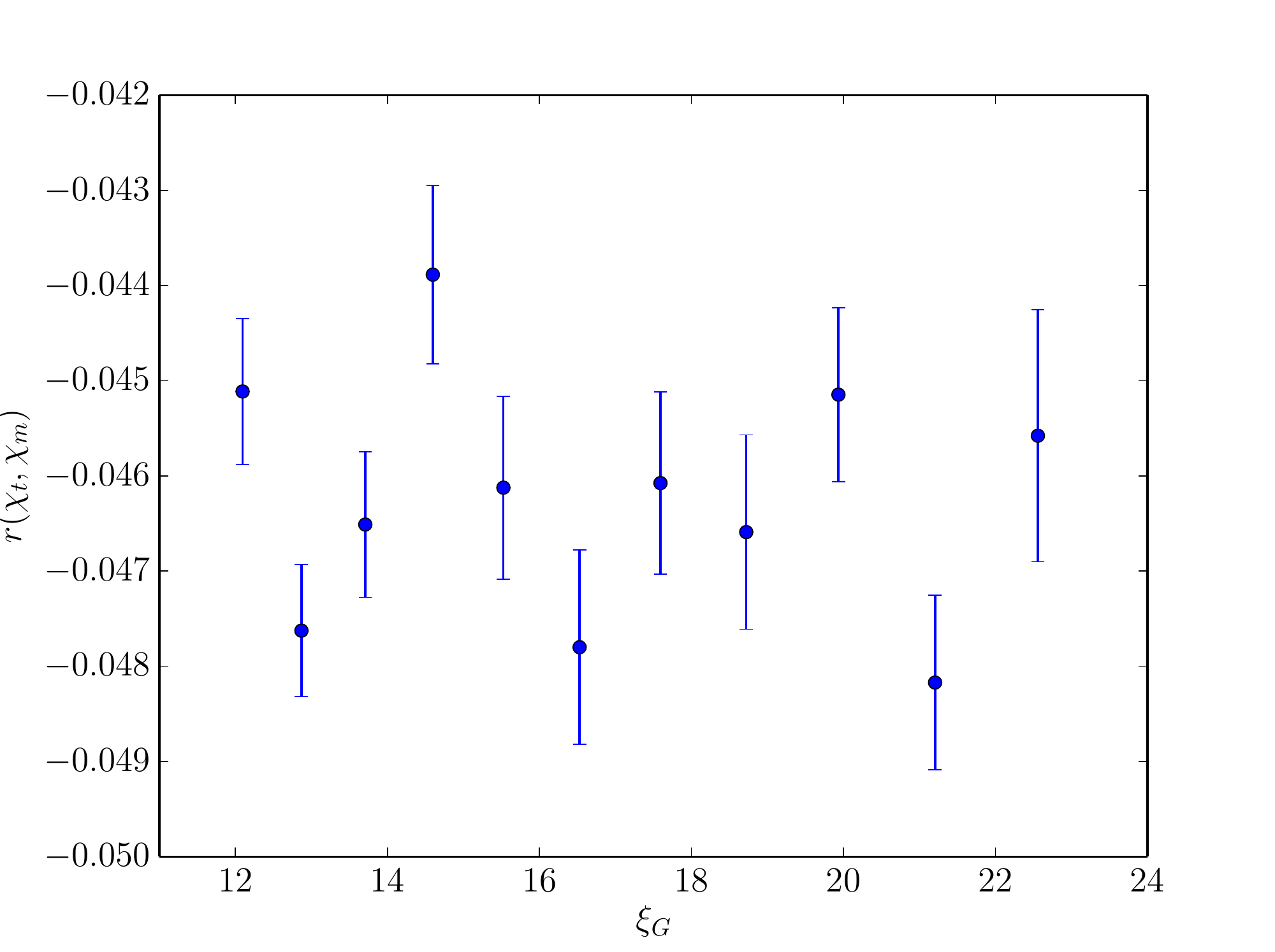} \\
	\vspace{-20pt}
\end{center}
	\caption{Plot showing the correlation coefficient $r$ between $\chi_t$ and $\chi_m$ for the highest values of $\xi_G$ measured. Errors were computed using a jackknife analysis.}
	\label{fig:correlation}
\end{figure}

However there is a very clear deviation from the expected power law scaling for $\chi_m$. We suspect that this is due to a small coupling to the topological modes. This idea was suggested in Ref.~\cite{HMC_CP9_results}, although the deviations from power scaling observed in this study could not be distinguished from momentum cutoff effects. Here however we take care to fit starting from much higher values of $\xi_G$; furthermore we explicitly verified that also including leading-order cutoff effects, i.e. powers of $1/\xi_G^2$ in Eqs. (\ref{eq:gaussian_scaling}) and (\ref{eq:top_scaling}), did not significantly change our results. As a simple test of the coupling between the two observables, we compute the correlation coefficient between $\chi_t$ and $\chi_m$ using
\begin{equation}
	r\left(\chi_t,\chi_m\right)=\dfrac{\mathrm{Cov}\left(\chi_{t},\chi_{m}\right)}{\sqrt{\mathrm{Var}\left(\chi_{t}\right)\mathrm{Var}\left(\chi_{m}\right)}}.
\end{equation}
The results of this analysis is shown in Fig.~\ref{fig:correlation}. We see very clearly that there is a small but highly significant correlation between the two. Ultimately we are left to consider an alternative ansatz for the scaling of the magnetic susceptibility.

\begin{figure}
\begin{center}
	\includegraphics[width=9cm]{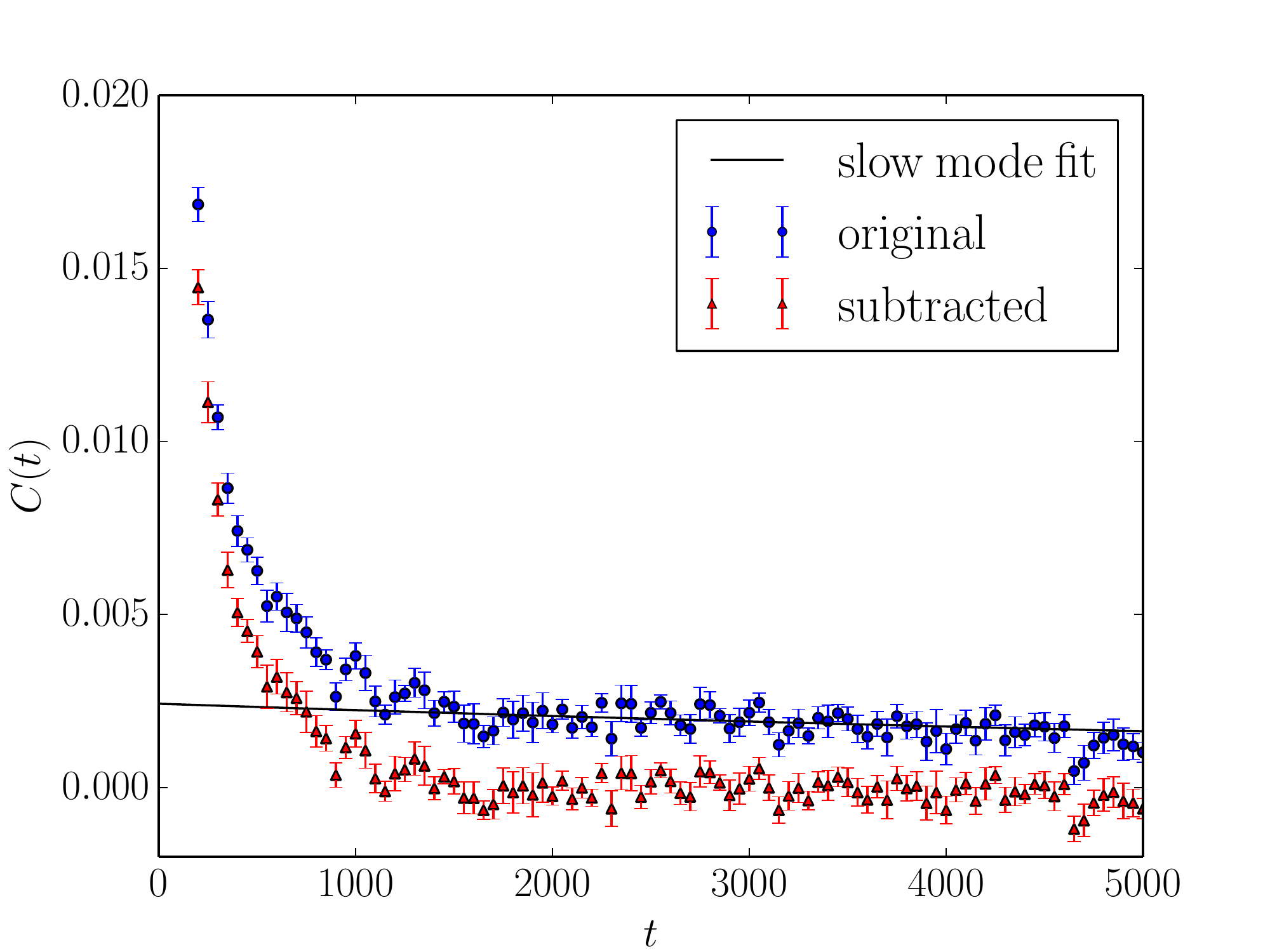}
	\vspace{-20pt}
\end{center}
	\caption{Plots of the autocorrelation function for $\beta=1.0$ data, sampled every 50$\mathrm{^{th}}$ timeslice, before and after the subtraction of the slow modes. The exponential decay of the slow modes is indicated by the best fit line.}
	\label{}
\end{figure}

\begin{figure}
\begin{center}
	\includegraphics[width=9cm]{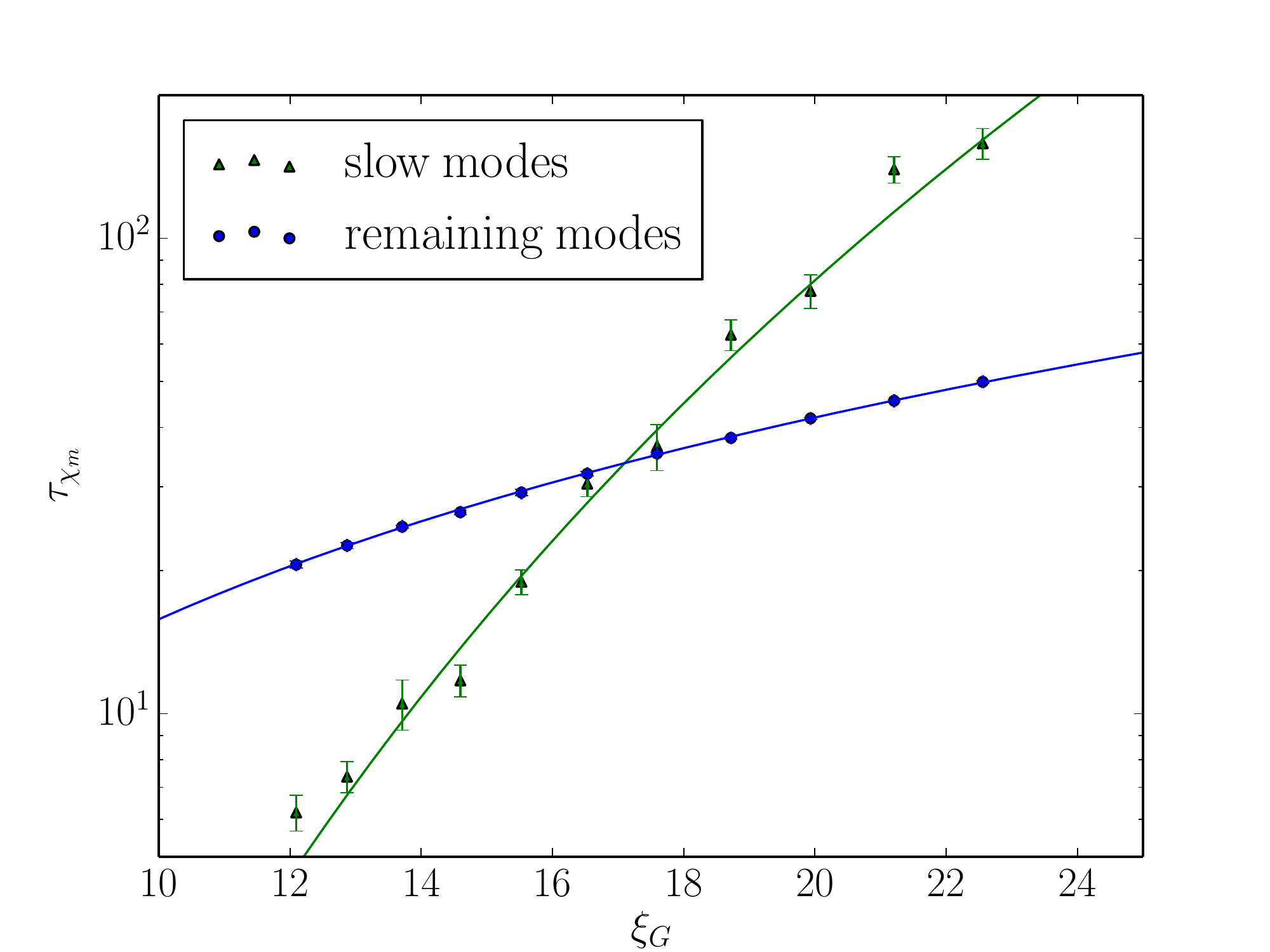} \\
	\vspace{-20pt}
\end{center}
	\caption{Plots of the scaling of $\tau_{\chi_m}$ with $\xi_G$ for both the slow modes and the remaining modes after the slow modes have been isolated.}
	\label{fig:tail_vs_no_tail}
\end{figure}

As originally noted in Ref.~\cite{HMC_CP9_results}, the autocorrelation function for the magnetic susceptibility displays a long, slow decaying exponential tail that contributes significantly to the integrated autocorrelation time at high values of the correlation length. We therefore separated the integrated autocorrelation time into two parts. First we take the contribution of the slow modes, which we estimate by fitting the tail of the autocorrelation function to a single exponential decay in a region $t\gg\tau$. We were careful to fit the tail in a region such that the fit is stable with respect to the time at which the fit begins. We then subtract this single exponential mode from the entire autocorrelation function and then recompute $\tau_{\chi_m}$ using the standard windowing procedure. In the Appendix we provide a brief proof of the validity of this method. Our expectation is that the slow modes should exhibit the same scaling behaviour as the toplogical modes. We display plots of the scaling trends for the slow modes and the remaining modes in Fig. \ref{fig:tail_vs_no_tail}. Fitting on $\xi_G>12.8$, we can immediately make the following observations: firstly the scaling of the integrated autocorrelation time of the remaining modes satisfies the expected power law scaling, with $\chi^2/\mathrm{dof}\simeq0.58$ and $p=0.80$. We obtain the fitted parameters $a=0.61(2)$ and $b=1.41(1)$; the exponent $b$ is consistent with quasi-Gaussian scaling~\cite{CSD_Topological_modes}. Secondly, the contribution of the slow modes overtakes the contribution of the remaining modes at around $\xi_G\simeq17.1$. Lastly, if we fit the slow modes to a power law fit we obtain an acceptable fit to the data with $\chi^2/\mathrm{dof}\simeq1.66$, $p=0.11$, $a=3.6(1.6)\times10^{-6}$ and $b=5.66(15)$. Crucially we notice that the exponent $b$ of the power law scaling of this mode is in strong agreement with the power law scaling exponent of the topological modes ($b=5.56(6)$); the low value for $a$ is consistent with the claim that the coupling to the topological modes is rather weak. An exponential fit for the slow modes also gives a fit that is narrowly acceptable, with  $\chi^2/\mathrm{dof}\simeq1.93$, $p=0.07$, although the errors on the fitted parameters are too large to draw any meaningful comparisons. Finally as a consitency check we confirmed that adding the integrated autocorrelation times of the slow modes and the remaining modes reproduce results that are compatible with our originally measured values. Altogether this analysis suggests that a reasonable ansatz for the scaling of the magnetic susceptibility is a double power law fit of the form
\begin{equation}
	\tau\sim a\xi^b+c\xi^d
\end{equation}
(or equivalently the sum of a power law and exponential fit).

\begin{figure}
\begin{center}
	\includegraphics[width=9cm]{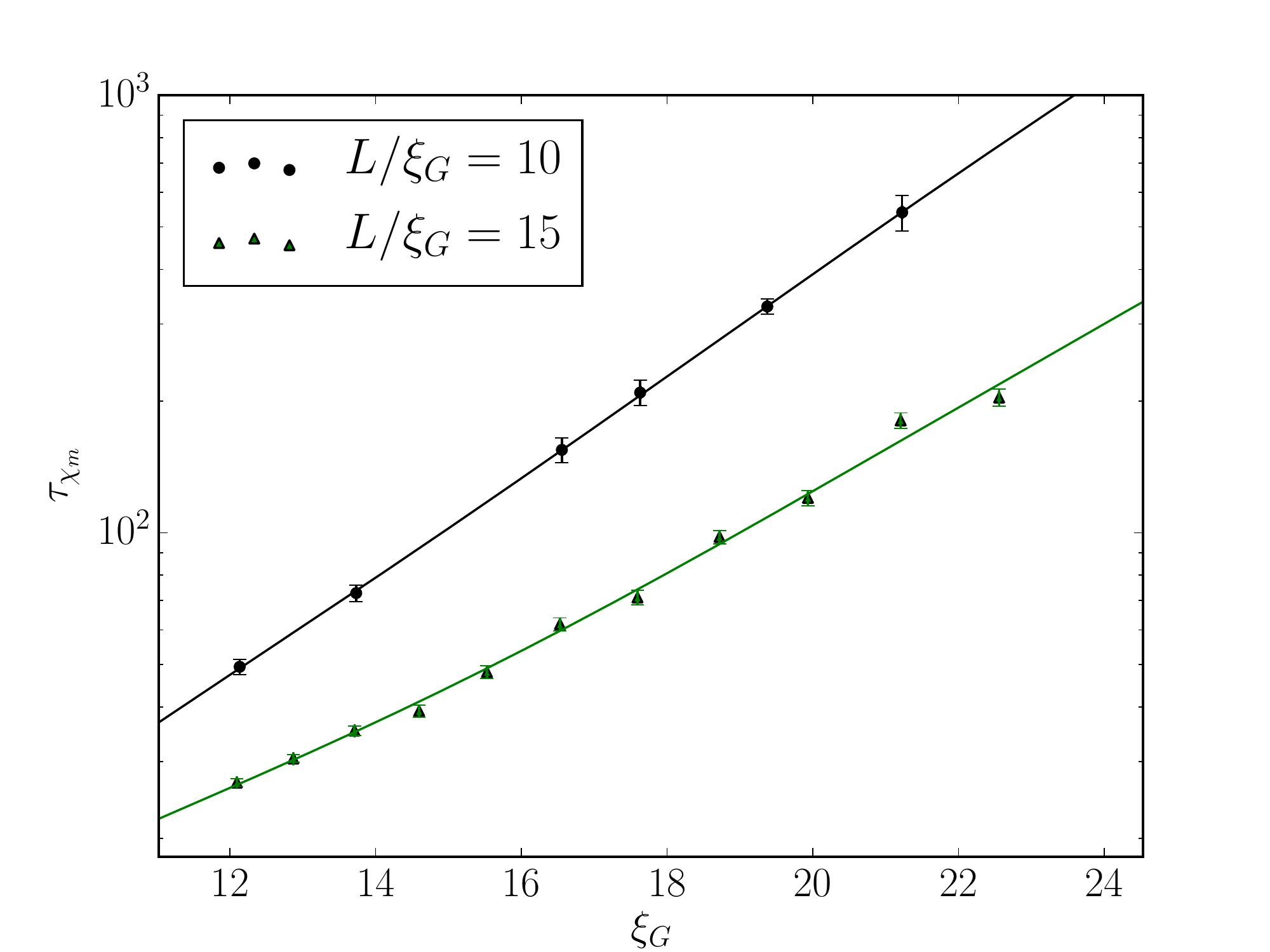} \\
	\vspace{-20pt}
\end{center}
	\caption{Plots of the scaling of $\tau_{\chi_m}$ with $\xi_G$ for our results with $L/\xi_G\simeq10$ and $L/\xi_G\simeq15$. The scaling is fitted to a double power law ansatz.}
	\label{fig:old_vs_new}
\end{figure}

\begin{figure}
\begin{center}
	\includegraphics[width=9cm]{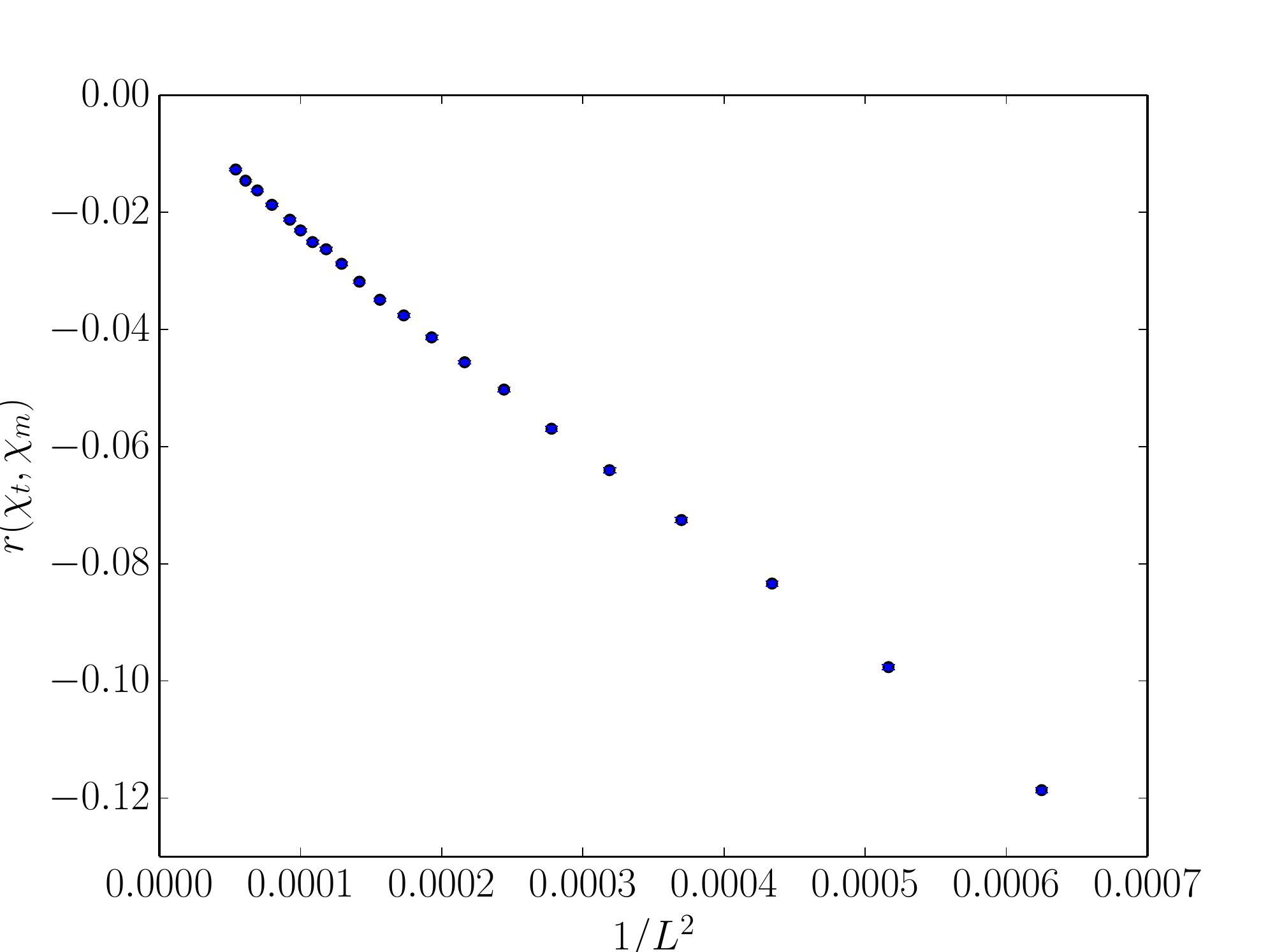} \\
	\vspace{-20pt}
\end{center}
	\caption{Finite volume scaling of the correlation coefficient between $\chi_t$ and $\chi_m$ in the CP$^9$ model at $\beta=0.8$.}
	\label{fig:coeff_FSS}
\end{figure}

To corroborate our results we also ran a set of simulations with $L/\xi_G\simeq10$. In Fig.~\ref{fig:old_vs_new} we display a plot of our results for the scaling of $\tau_{\chi_m}$. We remark that the scaling of the integrated autocorrelation time again deviates from the expected power law trend, and is even more severe in this smaller volume. We attribute this to the fact that the coupling to the topological sector appears to be a fairly strong finite volume effect, and thus varies significantly as $L/\xi_G$ changes. We remark that the finite volume dependence of $\tau_{\chi_m}$ was measured at a low value of $\beta$ where quasi-Gaussian scaling is dominant. At this value of $\beta$ the slow modes that couple to the topological sector can not be isolated and extracted as explained above and thus we do not have sufficient information to be able to account for finite volume effects in the scaling of the slow modes. This may account for the observed small deviations from the expected trend, particularly for large values of $\xi_G$ where the slow modes provide the majority of the contribution to $\tau_{\chi_m}$. However we can still measure the correlation between $\chi_t$ and $\chi_m$ for this data set; a plot of which is shown in Fig~\ref{fig:coeff_FSS}. This plot demonstrates that the scaling of the correlation coefficient towards the infinite volume limits appears to be dominantly power-like. Importantly we note that for the volumes we can feasibly simulate the correlation is non-negligible and thus could introduce a systematic bias on $\chi_m$ were the topological sectors not adequately sampled.

\section{Conclusions}

Through our high-statistics simulations of the CP$^9$ model we can draw several conclusions. Firstly, we have determined that for the level of precision reached in our simulations, taking $L/\xi_G\gtrsim15$ in the measurement of the topological and magnetic susceptibilities is sufficient in order to be able to neglect finite volume effects. However for $\xi_G$ we can resolve the finite volume effects up to $L/\xi_G\simeq25$.

Secondly we have shown that the CSD of the magnetic susceptibility is exacerbated by a small coupling to the topological sector. This results in a deviation from the expected power law form of CSD towards the character of the CSD of topological modes. Our simulations indicate that this dependence is a finite volume effect, as the correlation between topological observables and other observables (and by extension the CSD of these other observables) decreases with increasing physical volume. Nevertheless it follows that it is necessary to properly sample the topological sectors to avoid introducing systematic biases in the measurement of certain observables. Importantly there is no reason for such a feature to be absent in QCD simulations; in particular this may have significant implications for simulations which are frozen to a single topological sector.

\label{sec:Conclusions}

\section*{Acknowledgements}

The authors would like to thank Edwin Lizarazo for his contribution in the early stages of the project. We would also like to thank Luigi Del Debbio, Guido Martinelli and Ettore Vicari for interesting and useful discussions. We acknowledge the use of the ARCHER UK National Supercomputing Service (http://www.archer.ac.uk) in this work, as well as the IRIDIS High Performance Computing Facility, and its associated support services at the University of Southampton. The research leading to these results has received funding from the European Reasearch Council under the European Union's Seventh Framework Progamme (FP7/2007-2013) / ERC Grant agreement 279757. This work was supported by an EPSRC Doctoral Training Centre grant (EP/G03690X/1) and by the UK Science and Technologies Facilities Council (STFC) grant ST/L000296/1.

\appendix

\section{Analysis of autocorrelation functions}
\label{sec:auto_func}

In this section we present a theoretical description behind the subtraction of the slow modes of the autocorrelation function for the magnetic susceptibility. We assume that $\chi_m$ decomposes into two parts,
\begin{equation}
	\chi_{m}=\chi_{m}^{\prime}+c\chi_{t},
\end{equation}
which follows from the correlation we measured between these obserables. The quantity $\chi_{m}^{\prime}$ represents the contribution of modes which are not correlated with $\chi_t$.

We insert this expression into the formula for the autocorrelation function Eq. (\ref{eq:auto_func}), thus obtaining
\begin{align}
	C_{\chi_m}(t)=C_{\chi_m}(t) + c^{2}C_{\chi_t}(t) + 2\mathrm{Cov}_{\chi_{m},\chi_{t}}(t),
\end{align}
where $\mathrm{Cov}_{\chi_{m},\chi_{t}}(t)$ represents the cross-correlation between the observables. We neglect any remaining cross-correlation between these two observables, i.e. we take $\mathrm{Cov}_{\chi_{m},\chi_{t}}(t)=0$.

Lastly we assume that the autocorrelation functions display exponential asymptotic behaviour:
\begin{align}
	C_{\chi_{m}^{\prime}}(t)&\sim a_{1} \exp\left(-\dfrac{b_1}{\tau_{\chi_{m}^{\prime}}}t\right) \\
	C_{\chi_t}(t)&\sim a_{2}\exp\left(-\dfrac{b_2}{\tau_{\chi_t}}t\right).
\end{align}
It follows therefore that when we have $b_1/\tau_{\chi_{m}^{\prime}}\gg b_2/\tau_{\chi_{t}}$ the exponential behaviour of $C_{\chi_t}(t)$ will provide the dominant asymptotic behaviour of $C_{\chi_m}(t)$. At large enough values of $\beta$ this inequality will be satisfied; fitting the tail of the autocorrelation function at asymptotic times will thus yield the contribution of the slow modes to the integrated autocorrelation time.

\section*{References}
\bibliography{CPN}
\bibliographystyle{elsarticle-num}

\end{document}